# A Distributed Protocol for Detection of Packet Dropping Attack in Mobile Ad Hoc Networks

Jaydip Sen, M. Girish Chandra, P. Balamuralidhar, Harihara S.G., Harish Reddy

Embedded Systems Research Group, Tata Consultancy Services, Bangalore-560066, India

Email: {jaydip.sen, m.gchandra, balamurali.p, harihara.g, h.reddy}@tcs.com

*Abstract*—**In multi-hop mobile ad hoc networks (MANETs), mobile nodes cooperate with each other without using any infrastructure such as access points or base stations. Security remains a major challenge for these networks due to their features of open medium, dynamically changing topologies, reliance on cooperative algorithms, absence of centralized monitoring points, and lack of clear lines of defense. Among the various attacks to which MANETs are vulnerable, malicious packet dropping attack is very common where a malicious node can partially degrade or completely disrupt communication in the network by consistently dropping packets. In this paper, a mechanism for detection of packet dropping attack is presented based on cooperative participation of the nodes in a MANET. The redundancy of routing information in an ad hoc network is utilized to make the scheme robust so that it works effectively even in presence of transient network partitioning and Byzantine failure of nodes. The proposed scheme is fully cooperative and thus more secure as the vulnerabilities of any election algorithm used for choosing a subset of nodes for cooperation are absent. Simulation results show the effectiveness of the protocol.**

## I. INTRODUCTION

In a wireless ad-hoc network, a collection of mobile devices (referred to as 'nodes') with wireless network interfaces form a temporary network without the aid of any fixed infrastructure or centralized administration. MANETs have some special characteristic features such as: (i) unreliable wireless links used for communication, (ii) constantly changing network topologies and memberships of nodes, (iii) limited bandwidth of the links, (iv) Low battery lifetime, (V) Limited computation power of the nodes that prohibit the deployment of complex routing protocols and encryption algorithms for security. While these features are essential for flexibility and adaptability of various operations in MANETs, they introduce specific security concerns like vulnerabilities to link attacks including passive eavesdropping, active interfering, leakage of secret information, data tampering, impersonation, message-replay, message distortion and denial-of-service. An additional problem in MANETs is the security vulnerability of the routing protocols. A set of nodes in a MANET may be compromised in such a way that it may not be possible to detect their malicious behavior easily. Such nodes can generate new routing messages to advertise non-existent links, provide incorrect link state information, and flood other nodes with routing traffic thus inflicting Byzantine failure in the network. Another common routing disruption attack in MANETs has been packet-dropping attack by a group of malicious nodes. A group of nodes acting in collaboration may drop packets in the network at such a rate that the message communication in the network may be severely degraded and sometimes even completely disrupted. The detection of these malicious nodes may not be easy as they work in a group. Although there has been lot of research on detection and prevention of such an attack in MANETs, most of these schemes have either low detection rate, high complexity of detection algorithms, security vulnerabilities in the schemes themselves or high rate of false positives (Section II). In this paper, a mechanism for detection of malicious packet dropping attack in MANETs has been presented. The scheme involves a collaborative distributed protocol that utilizes complementary relationship between cryptographic key distribution and intrusion detection activity for detection of malicious packet dropping attack. The scheme has been evaluated for its performance by implementing it on the network simulator *ns-2*. The effectiveness and efficiency of the proposed mechanism has been compared with one of the currently available schemes and is found to have produced better results.

The rest of the paper is organized as follows. Section II presents some related work on defense against packet dropping attack in MANETs. Section III discusses the details of the proposed security mechanism. Section IV presents the simulations conducted on the mechanism and analyses the results. Section V concludes the paper.

## II. RELATED WORK

A number of works have been done on the area of ad hoc network security especially for detection of packet dropping attacks by malicious nodes. This section mentions some of these works.

To solve the problem of reduction in the throughput due to selfish and malicious nodes in a MANET, Marti et al [1] proposed two additional components to the dynamic source routing protocol (DSR): *watchdog* and *pathrater*. When a node forwards a packet, the node's watchdog verifies whether the next node in the path also forwards the same. The watchdog does this by listening promiscuously to the next node's transmissions. If the next node does not forward the packet, then it is misbehaving. The pathrater assesses the results of the watchdog and selects the most reliable path for packet delivery. However, this scheme has several drawbacks. First of all, overhearing does not always work particularly in situations like collisions or weak signals. Secondly, pathrater actually does not punish malicious nodes that do not cooperate in routing. Rather it relieves them of the burden of forwarding packets for others, while their messages are forwarded without any problem. In this way, the malicious nodes are rewarded and reinforced in their behavior.

CONFIDANT [2] protocol as proposed by Buchegger et al extends the concepts of watchdog and pathrater. In this mechanism, misbehaving nodes are not only excluded from forwarding route replies, but also from sending their own route request. The scheme includes a trust manager to evaluate the level of trust of alert reports and a reputation system to rate each node. The reports from trusted sources are only processed by the nodes. However, it is not clear how fast the trust level can be adjusted for a compromised node especially if it has a high trust level initially.

Buttyan et al [3] have advocated the use of tamper-resistant hardware on each node of a MANET to encourage cooperation. Nodes are assumed to be unwilling to forward packets unless they are stimulated to do so. In this approach, a protected credit counter runs on the tamper-resistant device. It increments by one when a packet is forwarded. It refuses to send its own packets if the counter is smaller than a threshold. Public key cryptography is used to exchange credit counter information among the neighbors and verify if forwarding is really successful. However, the availability of tamper-resistant hardware is a very vital assumption for the successful working of the scheme that involves complexity in hardware design.

In [4], the authors have presented a security architecture for MANETs involving mobile agents. In this scheme, multiple sensors deployed throughout the network collect and merge audit data implementing a cooperative detection algorithm. Sensors are deployed on some of the hosts in the network that monitor the network traffic. The selection of these nodes is based on their *connectivity index* and a distributed voting algorithm. The detection decisions are taken by mobile agents that migrate their execution and state information between the different sensor hosts of the network, and finally return to the originator host with the results. The authors have proposed two different methods of decision-making: collaborative and independent. They argue that independent decision-making by mobile agents is susceptible to single point of failure problems, and therefore the collaborative method should be used. The main advantage of this scheme is the restriction of computation-intensive operations of the system to few dynamically elected nodes. However, most of the available mobile agent frameworks are heavyweight and can often be the targets of attacks themselves [5].

## III. THE PROPOSED FRAMEWORK

This section presents the details of the proposed scheme. At first, some salient features of the scheme are described and then the details of the framework and the associated protocols are presented.

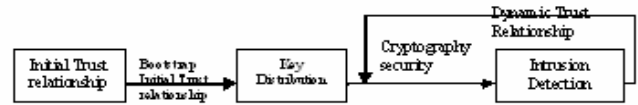

Fig.1. Key distribution and intrusion detection as complementary functions

### A. Salient Features of the Proposed Scheme

The proposed framework employs the complementary relationship between key distribution and intrusion detection. Key distribution in an ad hoc network require a trust management scheme to dynamically bind trust relationships between the key distribution servers and the clients. Usually, the context of this trust relationship is whether the node is well behaving or not. This requirement of dynamic trust management scheme can be satisfied by an intrusion detection system (IDS) that monitors the behavior of the nodes in the network for identification of malicious or faulty nodes (Fig. 1). The intrusion detection system, in turn, requires the security provided by the key distribution process through cryptographic techniques. This complementary relationship between key distribution and intrusion detection has been deployed in the proposed scheme to provide a high level of robustness into it. In the bootstrapping phase, the system uses the initial trust relationships that may be implemented by *location limited side channels* (LLCs) [7]. This provides the initial security to the intrusion detection mechanism, which in turn provides a dynamic trust management scheme for key distribution.

Due to dynamic nature of ad hoc networks, any intrusion detection (in this present context, detection of packet dropping) process should involve a distributed and cooperative protocol among the participating nodes. The cooperation between the nodes may be restricted within a small subset of nodes that are believed to be more trustworthy, or it may involve all the nodes in the network. Unlike most of the existing approaches, the proposed mechanism involves all the nodes in the network for working of the distributed protocol because the protocol involving a subset of nodes have the following drawbacks. Firstly, these schemes require some mechanisms to dynamically identify the subset of nodes that will participate in the protocol execution. Moreover, such schemes fail to take into account the observations of all the nodes in the network for identification of occurrences of events, and depend on the observations made by the nodes belonging to the subset only. For example, if the neighbors of a suspicious node cooperate to detect whether that node is really malicious, then the neighbors do not have information about the past behavior of the node as observed by other nodes in the network because of the dynamic nature of the network's topology. This may lead to incorrect evaluation of the behavior of the suspicious node.

A detection mechanism for malicious packet dropping attack that is based on a cooperative algorithm may be susceptible to attacks by Byzantine nodes. These nodes may make false claims of detecting malicious activities by some nodes that are really honest. The proposed scheme is secure and will operate correctly even in the presence of such Byzantine nodes in the network.

As in a MANET, every node acts as a router and participates in packet forwarding, there is lot of redundancy of routing information in the network. This redundancy of routing has been utilized in the proposed scheme to achieve a high degree robustness in its functioning so that it can work correctly in presence of selective packet dropping, packet tampering and even in the scenario of transient network partitioning.

### B. The Proposed Security Protocol

In the proposed scheme, every node in the network monitors the behavior of its neighbors and upon detecting any abnormal action by any of its neighbors invokes a distributed algorithm to ascertain whether the node behaving abnormally is indeed malicious. The protocol works through cooperation of some security components that are present in each node in the networks. These components are as follows: (i) monitor, (ii) trust collector, (iii) trust manager, (iv) trust propagator, (v) whistle blower. The functions of these components are described below.

(i) *Monitor*: The monitor module of each node passively listens to the communication to and from each of its neighbors. For detecting packet drops and modifications by the neighboring nodes, the monitor module of a node randomly copies the incoming packets to its neighbors and checks whether the neighbors really forward the packets with contents unchanged, or drop them, or modify the contents before forwarding them. The collected data is audited by the monitor. The deviation from normal behavior of a neighbor is used as an indicator for the unbiased degree of maliciousness, because this is independent of the past behavior of the neighbor node. If this unbiased deviation exceeds a pre-set threshold, the trust collector module of the node is invoked.

(ii) *Trust collector*: The Trust collector module of a node invokes a majority consensus algorithm among the neighbors of a node that has been suspected to be malicious. On being activated by its monitor module, the (accuser) node that has suspected some malicious activity by one of its neighbors challenges the suspicious node to verify its behavior as observed by all of its neighbors (Fig. 2). The accused (suspected) node on receiving the challenge responds by acknowledging the message and sending a *verify_behavior* message to all of its neighbors. The neighbors respond by sending the observed value of the degree of maliciousness of the accused node. The accused node calculates the group's trust in its behavior using the received values and broadcasts the computed group-trust along with the received responses to all the neighbors. All the messages are cryptographically secured by public key cryptographic mechanisms. The messages are also time-stamped so as to prevent replay attacks. For computing group trust value from the received responses, any consensus-based scheme can be used. In the proposed scheme, the difference of the absolute trust values and the average degree of maliciousness of the majority of the respondents (neighbors) has been taken as the final group-trust value of the node. Majority among the neighbors has been taken as the larger of the two subsets of nodes obtained by partitioning the nodes on the basis of a preset threshold value of trust.

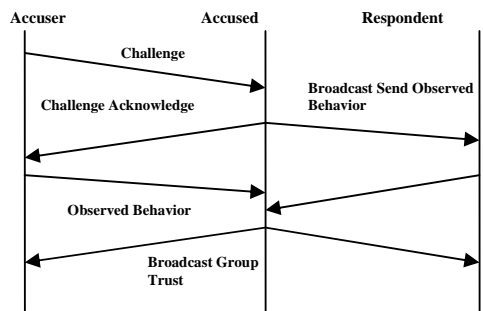

Fig.2. Challenge and response mechanism in Trust collector module

(iii) *Trust Manager*: Each node in the network maintains a global trust state containing the suspected nodes and their trust values. A table is also maintained that contains a list of nodes that has been determined to be malicious and thus should not be allowed any access to the network resources. The trust manager of a node is responsible for verifying the correctness of the group trust certificate received, caching them, and updating the global trust state (table) of the node for which it has received a new group certificate (from the neighbors of a suspected node). While verifying the correctness, the trust manager must check whether the response from every neighbor node has been correctly considered in computing the group- trust by the suspected node, and the messages have not been tampered with. This is implemented by cryptographic mechanisms. The contribution of a trust certificate in the final trust value of a suspected node depends on the global trust state of the majority of the neighbors of that node. If the majority of the neighbors observe that node is behaving maliciously, i.e., its trust value is low, the received trust certificate is propagated to all the neighbors of the suspected node. If the computed trust value of a node falls below the threshold trust level, a global alarm is raised and the *whistle blower* module is called on.

For updating the trust value of a node, a cumulative function is used. In (1), $T_{old}$, $T_{new}$, $T_{certificate}$ stand for the old trust value, new trust value, and the group recommended trust value for a node respectively.

$$(1 - T_{new}) = \alpha(1 - T_{old}) + \beta(1 - T_{certificate}) - \delta \quad (1)$$

$\alpha$ and $\beta$ represent the weightage corresponding to the old trust value and the new trust value of the node respectively. $\delta$ is the trust replenishment factor over time. $\beta$ depends on three factors $\alpha_1$, $\alpha_2$, $\alpha_3$ and can be expressed as follows:

$$\beta = \alpha_1 \alpha_2 \alpha_3 \quad (2)$$

The parameter $\alpha_1$ is given by

$$\alpha_1 = \frac{\sum\limits_{majority} w_i t_i}{W} \quad (3)$$

where, $w_i$, $t_i$ are the weightage and the trust value respectively of a node belonging to the majority group of the neighbors of the accused node. $W$ is a factor that depends on the size of the network. The factor $\alpha_2$ represents the weightage given to the new trust value computed, and $\alpha_3$ is defined as follows:

$$\alpha_3 = \begin{cases} 1 & if\ k = 1 \\ 0 & if\ k > 1 \end{cases} \quad (4)$$

where $k$ is the number of certificates received from the same group of neighbors or a subset of it in some threshold time interval.

(iv) *Trust Propagator*: The reputation propagator module uses mobility of the nodes for propagating trust certificates. Whenever a new trust certificate is computed for a node, it is initially distributed to a subset of the neighbor nodes of the suspected node. The size of this subset will determine the effective convergence time of trust information among the nodes that are at present and in near future would be the neighbors of the suspected node. At regular intervals, the neighboring nodes in the network participate in dynamic exchange of certificates. While the suspected node moves through the network, every node in the network would receive the certificate through flooding or exchange mechanism. The number of hops required to be flooded are determined dynamically by making the neighbors of the suspected node send their neighborhood information along with the observed behavior of the suspected node. The certificates are piggybacked on routing packets and thus they involve no communication overhead. This flooding and exchange mechanism enables detection of tampering of packets and provides robustness against packet dropping attacks as a node can compare a certificate in its local cache with the copies of the same certificate in its neighboring nodes. This scheme is also robust against network partitioning as the trust states of all the suspected nodes are maintained locally by all the nodes in the network. Moreover, due to group certification scheme, the number of false alarm is also less. As the number of malicious nodes in a network is usually small, the number of trust state to be maintained in the nodes are also few. Thus the scheme involves a very low storage overhead.

(v) *Whistle Blower*: The whistle blower module initiates a response action on receiving a global alarm about a suspected node. When a global alarm is raised, the alarm message is flooded across the entire network followed by the invocation of a voting algorithm among the nodes that have recently interacted with the suspected node, and a final decision is arrived at about the course of action to be taken (i.e., whether to isolate the node as it has been detected to be truly malicious or to keep it under surveillance as its trust value is still above the threshold). Fig.3 depicts the interactions of different security modules.

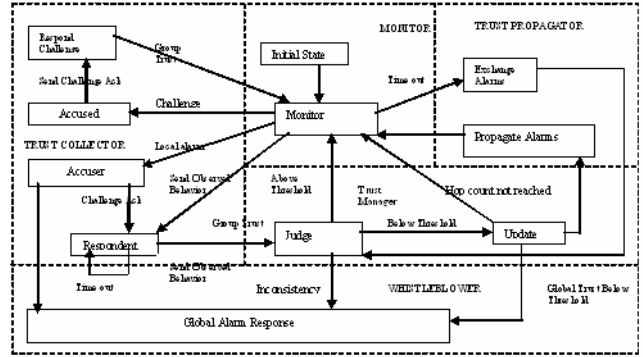

Fig. 3. Interaction among different security modules in a node

## C. The Packet Dropping Detection Algorithm

This section presents a formal method of detection of packet dropping attack used in the proposed scheme. The algorithm proposed in [1] has been extended. In the proposed algorithm, each node maintains a watch list of sent and overheard packets. Only those overheard packets are added to the watch list, which are destined to a neighbor node. Each sent packet is added to the watch list with a probability $p_1$ and each overheard packet is added to the watch list with a probability $p_2$. In Fig. 4, suppose node $A$ wants to send a packet to node $D$ via the intermediate nodes $B$ and $C$. $E$ overhears the transmission and both $A$ and $E$ check whether node $B$ really forwards the packet.

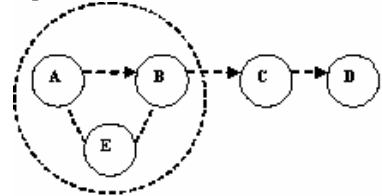

Fig. 4. Communication between different nodes in a MANET

However, there may be several reasons for which a sender node may not be able to overhear a packet sent. The reasons are: (i) the packet is maliciously dropped by the neighbor node (node $B$ in this example), (ii) the packet is dropped due to buffer overflow (congestion) at the forwarding node, (iii) the packet was forwarded by the forwarding node but there was a collision at the destination node (node $D$ in the example), (iv) the packet was forwarded after the preset timeout interval at the sender node. If $P_{malicious}$, $P_{congestion}$, $P_{collision}$, $P_{timeout}$ denote the probabilities of the above four events respectively, then we can arrive at the following equation:

$$P_{malicious} = 1 - \left( P_{congestion} + P_{collision} + P_{timeout} \right) \qquad (5)$$

Each node estimates the malicious drop probability using (5) and updates the probability of a node being malicious using the following equation:

$$P(n,t) = \alpha_1 P(n, t-1) + \alpha_2 f\left( packets dropped, P_{malicious} \right) \quad (6)$$

Where $n$ is the offending node $t$ is the time interval and $f\left( packets dropped, P_{malicious} \right)$ is a function of number of packets dropped and $P_{malicious}$. In the proposed mechanism, $f$ is chosen as an exponential function that initially rises slowly with the increase in number of dropped packets.

For evaluation $P_{malicious}$ in (5), $P_{congestion}$, $P_{collision}$ and $P_{timeout}$ have to be estimated. One approach of estimating $P_{congestion}$, at the forwarding node is to assume a Poisson arrival pattern [6]. In the proposed scheme, the congestion at the sending node is taken as the estimate of the congestion at the forwarding node. This seems to be a better estimate as all the nodes in the network are expected to have the same average traffic load on them. $P_{collision}$ is computed as the percentage of overlapping RTS (request to send) packets received. For example, suppose node $B$ sends a packet to node $C$ and overhears the medium for the forwarded packet. Suppose node $C$ really forwards the packet but at the same time node $E$ also sends a packet to $B$ resulting in a collision. Thus node $B$ will erroneously conclude that the packet is dropped by node $C$. $P_{collision}$ discounts computation of $P_{malicious}$ in such cases. This computation is based on the fact that node $B$ would have received RTS from both the nodes $C$ and $E$. If these RTSs have overlapping duration of bandwidth reservation, there is a chance that node $B$ will not be able to overhear the forwarded packet. $P_{timeout}$ accounts for the RTS collisions and the noise error in the medium and is estimated depending on the conditions of the wireless links.

## IV. SIMULATION AND RESULTS

### A. Simulation Environment

The proposed mechanism has been implemented in network simulator *ns-2*. Malicious behavior is simulated by dropping unicast packets at the network layer. In the simulation, all the links are assumed to be bi-directional. It is also assumed that promiscuous sniffing of packets is possible. This is, of course, true in 802.11 technology. The performance of the proposed mechanism is compared with that suggested in [1]. The parameters used in simulation are presented in Table I.

### B. Results

The performance of the proposed algorithm in terms of false alarm rate and successful detection rate has been compared with the *watchdog* algorithm proposed in [1]. In the simulation, each of the packets sent by a node is added to its watch list and each packet overheard by a node is put into its watch list (i.e., the probabilities $p_1$ and $p_2$ (Section III C) are taken as 1.0). A node is assumed to be malicious if its $P_{malicious}$ value exceeds 0.6. Fig.5 compares the false alarm rates as produced by the two algorithms. It is observed that the proposed approach reduces the false alarm rate by 50% as compared to the scheme suggested in [1]. This improvement is attributed to the estimation of $P_{congestion}$ in the proposed mechanism.

TABLE I SIMULATION PARAMETERS

| Parameters | Values |
| --- | --- |
| Simulation duration | 1000 s |
| Simulation area | 1000 m * 1000 m |
| Number of mobile nodes | 50 |
| Transmission range | 50 m |
| Movement model | Random waypoint |
| Traffic type | CBR (UDP) |
| Total number of traffic flows | 9 |
| Bandwidth threshold | 1 packet/s |
| Data payload | 512 bytes/packet |
| Number of malicious nodes | 5 (10% of total) |
| Maximum speed of a node | 8 m/s |
| Host pause time | 5 s |

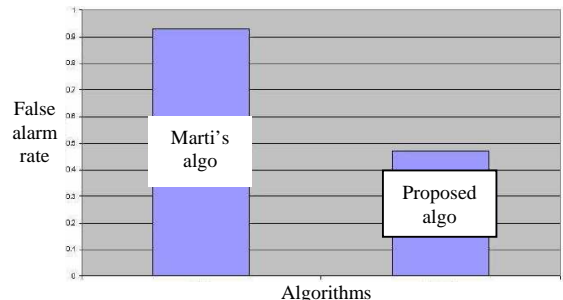

Fig. 5. Comparison of the false alarm rates

The comparison of the two algorithms in terms of successful detection rates is presented in Fig. 6. The decrease in the success rate of the proposed algorithm is due to over-estimation of congestion at a node that is really malicious and is dropping packets intentionally (i.e. not due to congestion). This simulation being a random instance, not all the malicious nodes are on the active traffic path, and thus not detected by the proposed algorithm. However, to avoid this situation the proposed scheme uses a cumulative function that assigns suitable weights to the past information about the node as well. Thus if a node is pretty much localized and consistently drops packets, it will be detected by the proposed scheme.

Each of the 50 nodes in the network is assigned a unique integer identification number from (0,49). Fig.7 shows the number of nodes that identifies a node as malicious corresponding to each node in the network. The malicious nodes are shown in dark. It can be seen that number of complaints is more for nodes that have higher packet dropping rates. Some of the nodes that are not malicious are also wrongly identified. In fact, these are the nodes that are experiencing heavy congestion and thus dropping packets at high rates. Thus the scheme also helps in identifying nodes that have higher congestion. This helps in reducing the number of false alarm as the nodes can take a distributed approach in arriving at a consensus to identify the malicious nodes ignoring the nodes that are experiencing congestion.

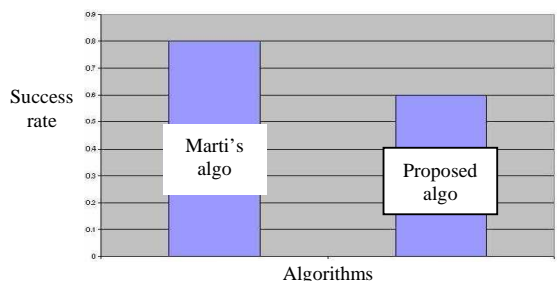

Fig. 6. Comparison of successful detection rates

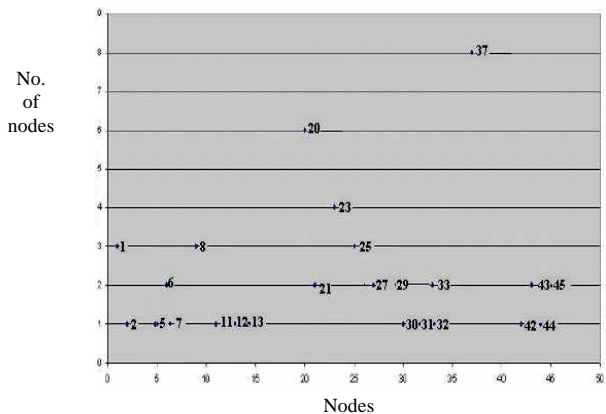

Fig.7. No. of nodes that finds a node malicious

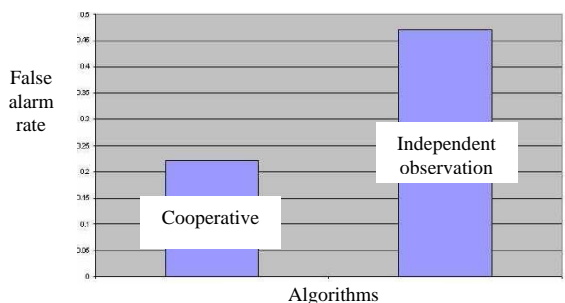

Fig.8. False alarm rates for distributed and non-distributed schemes

The effectiveness of a distributed consensus based approach in detection of malicious nodes is further depicted in Fig. 8 and Fig. 9. Fig. 8 shows the comparison of a distributed algorithm and an algorithm based on individual observation of the nodes. The distributed and cooperative approach reduces the false alarm rate by 50% as observed in Fig 8. Fig. 9 shows that in terms of successful detection rate both the distributed and individual-observation based algorithms have the same level of performance. The results thus clearly demonstrate the effectiveness of the distributed collaborative algorithm for detection of packet dropping attack in an ad hoc network.

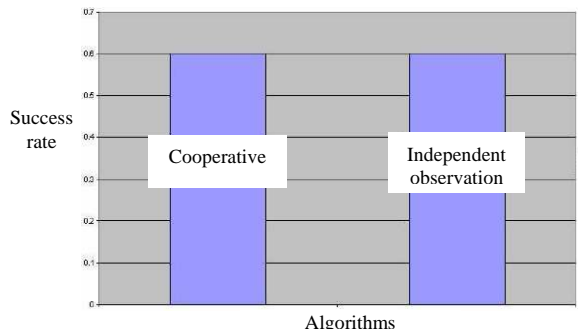

Fig.9. Success rates for distributed and non-distributed schemes

## V. CONCLUSION

In this paper, a distributed algorithm is presented for detecting malicious packet dropping attack in MANETs. The algorithm works on cooperative participation of all the nodes in the network at the network-bootstrapping phase but effectively identifies the nodes that behave maliciously as they participate in network activities. The redundancies in routing information in a MANET are suitably utilized to make the detection scheme highly robust and secure against various attacks. Due to the use of controlled flooding technique the mechanism has also very low communication overhead. Simulation carried on the scheme demonstrates its effectiveness. As a future scope of work, the mechanism can be extended so that the identified malicious nodes are isolated from the network and a secure routing protocol can be developed utilizing only the trusted nodes in the network.